# Cloud Adoption A Modern Approach

## Subhadip Kumar*

Western Governors University, skuma19@wgu.edu

Today's Information Technology world is cloud-centric. Companies are intrigued to migrate their workload private cloud from on-premise Datacenter to Public cloud to take advantage of the latest innovations. It drives the business growth and competitiveness of the organization. At the same time, Enterprise Architects need to understand the drawbacks and challenges to migrate the workload to Cloud. This paper aims to identify the key factors to migrate the workload to the cloud. It also helps an organization to identify the candidate for cloud migration. An impulsive decision to move to the Cloud may be detrimental to an organization. Also, I will discuss one case study to see the benefits and disadvantages of cloud migration. This will help the organization maximize its ROI.odays Information Technology world is cloud centric. Companies are intrigued to migrate their workload private cloud from on-premises Datacenter to Public cloud to take advantage of the latest innovations. It drives business growth and competitiveness of the organization. At the same time, it is important for Enterprise Architects to understand the drawbacks and challenges to migrate the workload to cloud. This paper aims to identify the key factors to migrate the workload to cloud. It also helps an organization to identify the candidate for cloud migration. An impulsive decision to move to the Cloud may be detrimental for an organization. Also, I will discuss one case study to see the benefits and disadvantages for cloud migration. This will help the organization to maximize the ROI.

## 1 INTRODUCTION

No Question entire world is adopting cloud computing and cloud-native technologies. It helps businesses to adopt the cloud for better scalability, resiliency, availability, and faster deployment. Another major factor is cost optimization – There is a general perception that the cloud is cheap. I will take a deep dive to see the overall cost benefit. An organization that is currently running all of the workload on-premise must consider several factors to determine the overall cloud strategy. We will see key deciding factors in this study – Cost, Existing Footprint, Security, Disaster Recovery, and Skill and Evolving Technology. In midsize to large organizations, run several different business modules starting from legacy applications, web applications, and modern software packages like SAP, Salesforce, etc. Often these applications are tightly coupled to each other, and the overall performance of the application depends on the performance of each technology stack. Over the years a lot of customization was made on these applications to meet the business requirement.

Also, it is important to understand whether the current architecture is fault-tolerant and has disaster recovery capabilities. What is the accepted range of RTO and RPO? Whether current architecture meets the demand. What is the main complaint from the business? Whether cloud adoption will help to meet the demand?

I will try to cover all of these in this paper. The paper is divided into 3 sections – Key Factors, a Case Study, and a Summary.

---

* Place the footnote text for the author (if applicable) here.

## 2 KEY FACTORS

Let's take a deep dive into the different key factors in this section.

### 2.1 Cost

A major driving force for cloud adoption is cost. The sales team of the public cloud frequently promotes the idea that cloud migration can significantly reduce the organization's costs. Another important aspect all Cloud Service Providers (CSP's) highlight is pay-per-use – while that sounds very cost-effective. As stated in [1] Just taking legacy applications and moving them to the cloud— "lift-and-shift"—will not automatically yield the benefits that cloud infrastructure and systems can provide. Cloud vs on-premises is like leasing vs owning a car. If you lease a car and drive more than your mileage limit in a year and if you are planning to have that car for a longer period, it is always cost-effective to own the car. In real life most organization run applications that are monolithic and run on fixed hardware. Moving them to the cloud by lift-and-shift will not be beneficial and cost-effective in most cases. On the other side if you are running a reporting application in Oracle DB which only generates reports on demand, migrating that workload to a cloud PaaS service might be beneficial from a cost perspective.

Other factors to consider are a substantial upfront investment to set up backbones for cloud adoption and often a multiyear journey. Cost to connect the primary region to on-premises data center – labor cost, Network Edge services (Point of Presence) cost, landing zone cost.

What applications will be migrated – whether it is a lift and shift or rearchitecting? For example, whether migrating a custom application from on-premises hardware to the cloud (IaaS) or rearchitecting an existing database to a PaaS service like RDS.

Therefore, a feasibility study is important to determine whether cloud migration is financially and technically feasible. Once the candidate applications are identified for migration a detailed Return of Investment (ROI) and Total Cost of Ownership (TCO) assessment needs to be carried out [13].

### 2.2 Existing Footprint

Another important factor to consider is the investment in existing infrastructure that includes building, power supply, server blades, virtualization licenses, OS licenses, storage, network, switches, routers, rack, etc. The organization needs to calculate existing running costs which include resources, and data center Opex costs. Whether rehosting the existing infrastructure to the cloud will have cost benefits or whether it can wait till the end of the lifecycle of the existing hardware. Some licenses are transferrable to the cloud and some or not. Therefore, cloud adaptations must be a phased approach and often land in a hybrid architecture. There are strict data residency requirements, there are audit requirements for certain services, and regulation requirements.

Financial organizations usually have a strong data residency requirement on consumer PII and financial data. That related application will always stay onsite. In such cases, organizations can leverage cloud technologies for internal applications such as employee payroll however consumer applications like banking or transaction software stay on-premises data centers.

Architects along with the legal team need to identify such applications whether can be migrated to the cloud and have benefits over current infrastructure when it is close to retire.

Sometimes it is a strategic direction that all new workloads will be cloud-native and the rest of the applications will be handled case by case basis. In such scenarios, organizations end up in a hybrid architecture and take advantage of existing investments as well as new technologies.

## 2.3  Security

To me, this is the most important factor to consider, and it is having several aspects. I'm trying to cover some aspects here.

### 2.3.1 Data Residency:

As mentioned in [2], data residency is still a major concern for any organization. Organizations are unaware of their sensitive data because the Cloud Service Providers (CSP's) maintain data centers in geographically distributed locations resulting in several security challenges and threats.

### 2.3.2 Service Level Agreement (SLA)

Service level agreement (SLA) is a contract between CSP and the Consumer (in most cases the organization) to define the maximum unavailability of the data. This is sometimes confusing and must be vetted carefully. Most of the time overall availability on an application is measured based on the availability of the individual component – such as storage, virtual machines, network etc. Whatever the lowest availability is the availability of the entire application. CSPs often use 9's showing the overall availability such as 99.9%, 99.95%, 99.99% etc. Figure 1 below represents the uptime graph. The organization needs to carefully review the current uptime with the CSP's uptime to see if it exceeds or meets the demand.

| Availability | 99% | 99.90% | 99.99% |
|---|---|---|---|
| Daily | 14m 24s | 1m 26.4s | 8.6s |
| Weekly | 1hr 40m 48s | 10m 4.8s | 1m 0.5s |
| Monthly | 7hr 18m 17.5s | 43m 49.7s | 4m 23s |
| Year | 3d 15hr 39m 29.5s | 8h 45m 57s | 52m 35.7s |

Figure 1: Uptime Graph

Also, keep in mind, sometimes it is hard to establish a case that a breach of SLA happened, and savings often come as prorated basis of the exceeded downtime. Even it can be established, saving usually comes as a form of credit for future consumption.

### 2.3.3 Multitenancy and Elasticity

There should be isolation to deliver secure multitenancy among tenant data. Also scaling up and down (elasticity) of consumer's resources gives possibility to other consumers to use previously assigned resources [3]. When the hardware doesn't provide adequate isolation, one malicious attack can impact other instances on the same instance [4].

### 2.3.4 Data Privacy and Encryption

As the data is residing in someone else data center data privacy and encryption is of utmost importance. Architects should construct a system to manage the encryption of data both in transit and in rest. It should also allow for the layering of encryption schemes and provide secure communication channels [5].

## 2.4   Disaster Recovery

The cloud offers a great extent of high availability and disaster recovery solutions both for IaaS as well as PaaS and SaaS solutions by leveraging their zonal and regional data centers. Therefore, organizations currently struggling to establish a robust business continuity solution cloud may be an answer. Finally, the determining factor is the RTO and RPO. "Cold" backup site comes with the lowest cost but with high RTO and RPO. "Warm" standby comes with moderate cost and moderate to low RTO and RPO whereas "hot" standby is the costliest and has lowest RTO and RPO. There are different types of disasters – natural, technical, human-made, rapid-onset, and slow-onset [7]– organizations must ensure that DR solutions can be able to handle all these scenarios.

The optimal disaster recovery planning should take into consideration the key parameters including the initial cost, the cost of data transfers, and the cost of data storage [8]. The DR solution must be able to detect a disaster and perform a failover procedure as well as revert the control back to the primary data center when the disaster is over. Physical DR on-premises site setup includes data center space, building, and additional operational costs like power, cooling, site maintenance, and staffing. However, in case the DR site exists pre-cloud era, these costs are minimum.

Figure 2 illustrates several DR options in cloud and its advantages and disadvantages.

| Scenario | Advantage | Disadvantage |
|---|---|---|
| When the primary application or data is in cloud and backup or recovery site is in private data centre. | Isolating two sites ensure availability when there is a global outage on Cloud | High ingress-egress cost. Additional networking/security cost. |
| When both primary site and recovery site are in cloud. | Highest flexibility - Take advantage of Zones and Regions, Lowest ingress-egress cost | If there is a global outage DR is impacted |
| When the application is in primary data centre and backup or recovery site is in cloud. | Low opex cost for "warm standby" | Additional networking/security setup is needed. |

Figure 2: Disaster Recover Option

Things to consider whether the organization is currently has a secondary site that can be used as a DR site or whether it can leverage the cloud as DR. Overall cost to standup a DR site, latency for end users, RTO and RPO.

## 2.5 Skill and Training

If an organization planning to adopt the cloud – they need to plan. With the introduction of cloud services, there is a need to define IT governance especially the shared responsibility model that defines the decision-making privileges of the individual and its impact to the organization. With the introduction of the cloud, there is an increasing need to either hire new resources with the required skillset or upskill the existing employees by providing training.

The executives and leaders within the organizations should be clear about the scope and understanding of the cloud and the primary motivation behind it [9]. As the best practice organization forms a Cloud Center of Excellence that handholds the employees during the transition phase.

Cloud adoption also requires behavioral change – it is a common human tendency to compare each cloud service with on-premise and that leads to some conflict in terms of governance. For example, in the cloud mostly it is DevOps or SecOps where developers have access to spin up VMs however in on-premises it is usually by the server team who provision it. Adaptation to culture shift is very important.

The leadership team plays a very important role once the direction is clear. Once the motivation and support are high by the leadership team cloud adoption is faster. When motivation is low there is skepticism among employees to put discretionary effort to upskill themselves. There is a lot of distrust in this journey, and it is the leadership team's responsibility to establish trust and clear any doubt among the employees. Once that phase is over, employees will be self-motivated and put effort to learn new skill sets.

Organizations can also hire new resources who can cross-train the employees to the cloud technologies. New resources will bring fresh ideas and enrich other employee's views toward cloud.

Usually, cloud providers promote some form of training and skill-building effort as part of the engagement that includes onsite training, online training, certification, etc. Management needs to ensure that they are freeing up resources from their existing duties so that they can fully focus on the training and maximize learning.

## 2.6 Evolving Technology

One of the main driving forces to migrate to the cloud is to take advantage of different microservices offered as PaaS or SaaS by the CSPs. Either existing monolithic legacy applications are unable to meet user needs or it is not flexible enough. These applications may be a good candidates for cloud migration. Microservices can be characterized by a number of principles such as evolutionary design, decentralized control of data, and design of failure [10]. This is known as function as a service – some well-known serverless solutions include AWS Lambda, Azure Functions etc. Many such microservices can be containerized [11]. Organizations often migrate such complex legacy solutions to the cloud.

The organization needs to check the effort it requires to perform the migration. Many such legacy applications over the years adopted a lot of customizations. Rearchitecting such applications require a considerable amount of planning and execution, and many times project managers use agile methodology to streamline the delivery.

Another major reason to migrate to the cloud is elasticity. Elasticity is the degree to which a system can adapt to workload changes by provisioning and de-provisioning resources in an automated fashion [12]. All CSPs provide a wide variety of elasticity which includes autoscaling when a certain threshold (like CPU/Memory) breaches a certain threshold. The autoscaling feature is available for both IaaS and SaaS applications. It also offers various caching technologies such as Content Delivery Networks (CDN), edge location, and regional caching.

Some of the object storage solutions offer 11 9's reliability and are extremely fault tolerant. Organizations can also host static websites on such storage without any further tools.

Utilization of such cloud-based services has tremendous cost benefits if it can be utilized properly.

## 3  CASE STUDY

In this case study I will explain how an organization can evaluate migrating on-premises SAP Applications to Cloud vs to another hardware on-premises.

An organization running a 6TB in-memory SAP HANA Application on-premises and planning to do a migration to the cloud as an IaaS solution. They are planning to develop a three-system landscape (Development, Quality, and Production) and the quality and production system will have high availability and disaster recovery capabilities. A detailed cost estimate study was conducted that includes a cost breakdown for 5,8 and 10 years, advantages and disadvantages, availability of skilled workforces, availability of cloud-native tools, published SLA, and product-specific migration tools.

Cost for 5 years was coming as $4.5M, 8 years $7.5M, and 10 years $10M, for similar infrastructure on-premises cost will be around $4.5M which means breakeven with the cloud is on the 5th year. Also in the cloud, organizations can start with a smaller capacity and later scale into a bigger VM – this way cost can be optimized for initial years.

While analyzing advantages – The cloud offers scalability – any time it requires more capacity it can be provisioned in days compared to on-premises where there are multiple stages and lead times are high – procurement, vendor negotiation, delivery, rack and stack, networking, and server build which takes more than 2 months. From a disadvantage point of view, SAP is a bandwidth-intensive software and when it talks to other tightly coupled on-premises systems, performance will be impacted.

Within the organization, upskilling is needed to perform installation and software maintenance with the cloud. All CSPs offer some initial funding programs to accelerate migration and training.

Availability of cloud-native tools is a big benefit – as for on-premises there are different vendors and there are several license costs that can all be consolidated. Management of the software is also easy by using cloud-native tools for backup, patching, monitoring, etc. so that we don't have to use third-party software for each activity.

SLA offered by the CSP is 99.95% which translates to monthly ~22m and yearly 4h 21m unplanned outage. When calculated on-premises unplanned outage it is very close to the SLA offered by the CSPs.

The cloud offers very detailed migration tools that can be leveraged to perform the migration in a timely fashion. Migration tools have been tested on their internal lab as well as various customers and optimized over time. On-premises to on-premises migration is easy and can be carried out on an internal corporate network. SAP vendor offers different tools to ease the migration.

Security which is another major deciding factor – on-premises is a clear winner as organizations have more granular control over the cloud as well as isolation however cloud security and isolation increased over time.

Cloud is always in an advantageous position because of the zonal and regional presence and organizations can leverage availability zone for high availability and regions for disaster recovery solutions.

Figure 3 illustrates a weighted matrix that can be used to determine whether migrating to the cloud is a good option. In this scenario, cloud migration is recommended.

| Success Criteria | Cloud Ranking (0 - 5) | On-Premises Ranking (0-5) | Weighted (1-5) | Cloud Score | OnPremises Score |
|---|---|---|---|---|---|
| Cost | 5 | 4 | 5 | 25 | 20 |
| Advantages/Disadvantages | 5 | 3 | 4 | 20 | 12 |
| Skilled Workforce | 3 | 5 | 4 | 12 | 20 |
| Cloud Native Tools | 4 | 2 | 3 | 12 | 6 |
| Published SLA | 4 | 4 | 4 | 16 | 16 |
| Migration Tools | 4 | 3 | 3 | 12 | 9 |
| Security | 3 | 5 | 5 | 15 | 25 |
| DR Capabilities | 5 | 4 | 4 | 20 | 16 |
| | | | | | |
| Total Score | | | | 132 | 124 |

Figure 3: Evaluation Criteria

## 4  SUMMARY

The objective of this paper is to remind organizations need to do sufficient due diligence to fully understand cloud computing and its pros and cons. Sometimes the issue is the mismatch of expectation between the client and the CSP and often leads to a hostile relationship which is detrimental to both parties.

## 5  ACKNOWLEDGMENTS

I would like to thank anonymous reviewers for the comments and suggestions.